\documentclass[useAMS,usenatbib]{mn2e} 
\usepackage{graphicx} 
\def\Lsun{L$_\odot$}  
 
\def\msun{M$_\odot$} 
\def\mass{$\mathcal{M}$}
\def\massun{$\mathcal{M}_\odot$}

\def\Mv{M$_{\rm v}$}  
  
\def\teff{T$_{\rm eff}$}  
\def\0BMV{$(B-V)_{\rm 0}$}

\def\simgt{\lower.5ex\hbox{$\; \buildrel > \over \sim \;$}}  
\def\simlt{\lower.5ex\hbox{$\; \buildrel < \over \sim \;$}} 
\title[85~Pegasi]{85~Peg A: which age for a low metallicity solar like star?} 
\author[D'Antona et al.]{F. D'Antona$^{1}$\thanks{E-mail: 
dantona@mporzio.astro.it},   
D. Cardini$^{2}$, M. P. Di Mauro$^{3}$, C. Maceroni$^{1}$, 
I. Mazzitelli$^{4}$  \newauthor and J. Montalb\'an$^{5}$
\\
$^{1}$INAF, Osservatorio Astronomico di Roma, via Frascati 33, I-00040 MontePorzio, Italy\\ 
$^{2}$INAF, Istituto di Astrofisica Spaziale, via del Fosso del Cavaliere 100, I-00133 Roma, Italy\\ 
$^{3}$INAF, Osservatorio Astrofisico di Catania, via S. Sofia 78, I-95123 Catania, Italy\\ 
$^{4}$via Zambarelli 22, I-00044 Frascati, Italy\\ 
$^{5}$Istitut d'Astrophysique et G\'{e}ophysique Universit\'{e} de Li\`{e}ge,
			All\'{e}e du 6 A\^{out}, B-4000 Li\`{e}ge, Belgium\\
}
\begin{document} 
 
 
\pagerange{\pageref{firstpage}--\pageref{lastpage}} \pubyear{2005} 
 
\maketitle 
 
\label{firstpage} 
 
\begin{abstract}
 
 We explore the possible evolutionary status of the primary component of the 
 binary 85~Pegasi, listed as a target for asteroseismic observations by the 
 MOST satellite. In spite of the assessed `subdwarf' status, and of the 
 accurate distance determination from the Hipparcos data, the uncertainties in 
 the metallicity and age, coupled with the uncertainty in the theoretical 
 models, lead to a range of predictions on the oscillation frequency spectrum. 
 Nevertheless, the determination of the ratio between the small separation in 
 frequency  modes, and the large separation as suggested by Roxburgh (2004), 
 provides a very good measure of the star age, quite independent of  the 
 metallicity in the assumed uncertainty range.  In this range, the constraint 
 on the dynamical mass  and the further constraint provided by the assumption 
 that the maximum age is 14~Gyr limit the mass of 85~Peg~A to the range from 
 0.75 to 0.82~\msun. This difference of a few hundreths of solar masses leads to 
 well  detectable differences both in the evolutionary stage (age) and in the 
 asteroseismic properties. We show that the age determination which will be 
 possible through the asteroseismic measurements for this star is independent 
 either from the convection model adopted or from the microscopic metal 
 diffusion. The latter conclusion is strengthened by the fact that, although 
 metal diffusion is still described in an approximate way, recent observations 
 suggest that the real stars suffer a smaller metal sedimentation with respect 
 to the models.
 \end{abstract} 
 
 
\section{Introduction} 
 
The bright visual binary 85~Peg (HD 224930) has been  the object of  numerous 
studies since the time of its discovery by \citet{Burham1879}. Spectroscopy 
revealed a metal deficient object and astrometry moderately high space velocity 
components \citep{APal04, nordstrom}, both suggesting a classification as 
subdwarf and a probable old age.  Its orbital elements are known with good 
approximation since a long time, and  recent determinations, made through very 
precise astrometric, photometric and  spectroscopic measurements \citep[][ and 
references therein]{Griff04}, have not  substantially changed what has been 
known for more than one century nor have been able  to solve the problem  of 
the masses of the stars that  constitute 85~Peg. Indeed, while it  seems 
evident that, making the reasonable hypothesis of a common origin, the 
secondary star should have a considerably smaller mass than the primary, due to 
the about three  magnitudes of difference, the masses determined dynamically 
are similar.  This discrepancy has  been ascribed to a third much fainter 
component \citep{Hall48}.   In this situation the determination of the two 
masses has been  committed up to now to theoretical calibrations by modeling 
stellar parameters  (Fernandes et al. 1998, Fernandes, Morel \& Lebreton 2002)
 on the basis of 
observational data and of assumptions on age, initial helium  content and 
metallicity. This leaves broad intervals of discretionality which,  coupled 
with the uncertainty in the theoretical models, lead to rather different 
determinations of mass and consequently of age. The determination of the age of 
subdwarfs sets a lower limit to the age of old galactic stars,  possibly 
significant as  lower limit for the age of the Galaxy.  

Recently, a new tool has become available to infer the evolutionary  status and 
the structural properties of a star, asteroseismology.  During the last years, 
numerous attempts have been made  with the aim of identifying oscillations in 
distant stars and also of modeling the stellar pulsational phenomena.  It is 
now clear, indeed,  that oscillations have several advantages over all the 
other observables: pulsational instability has been detected in stars in all 
the evolutionary stages and of different spectral type; frequencies of 
oscillations can be measured with high accuracy and depend in very simple way 
on the equilibrium structure of the model; different modes are spatially 
confined and probe different layers of the stellar interior. 
 
85~Peg~A can be considered as a good candidate for asteroseismic studies, since 
theory predicts a detectable surface amplitude of oscillations 
with a maximum value of $v_{\mathrm{osc}} \simeq 18.8$ cm~s$^{-1}$ according to the
scaling law of \cite{Kje95} 
and a mean value of $v_{\mathrm{osc}} \simeq 9$ cm~s$^{-1}$ (Houdek private communication)
obtained according to the theoretical simulation of modes
stochastically driven by convection \citep{houd}.
85~Peg~A, indeed,  is listed as  target for asteroseismic observations by MOST 
\citep{walker03},  the first satellite totally dedicated to the observation of 
pulsating  stars from  space.  
 
The aim of the present paper is to provide a  comprehensive theoretical study 
of the system. We will use new theoretical evolution  models  to find a 
correspondence between the  observed data and the derived  astrophysical 
parameters,  in order to locate  85~Peg~A in an  evolutionary context. We will 
discuss  the possibility of discriminating  among its possible structure models 
by means of the oscillation data which will be  soon available. 
 
A summary of known data on 85~Peg is presented in Section \ref{thorn}, the 
details of the evolutionary models and their result in Sections \ref{models} --
\ref{85peg},  the computations of the oscillation frequencies and a discussion 
of the results  in Section \ref{puls} and \ref{oscill}, which is followed by 
the final conclusions.  
 
\section{A "thorn in the side" of evolutionary theories: the visual  
binary 85~Pegasi}\label{thorn} 
 
The visual and single-lined spectroscopic binary 85~Peg is a bright (V=5.75), 
and nearby ($\sim 12$ pc) system with an orbital period of 26.31 yr 
\citep{Griff04}. Its small angular separation ($a=0\farcs 83$) and the marked 
magnitude difference between the components, $\Delta m \simeq 3.08 \pm 0.29$ 
mag in V \citep{tenBr00}, makes it a difficult target both for  visual and 
spectroscopic observations. The primary component is a metal deficient G5 
subdwarf \citep{Ful00}, while the most-of-the-time  unresolvable secondary is 
supposed to be, on the basis of the $\Delta m$ value, a K6-K8 dwarf.  The 
classification as subdwarf is based both on the space  velocity component 
values  and on its metallicity \citep[][ respectively $\mathrm{[Fe/H]}=-0.88$ 
and $-0.80$]{APal04, nordstrom}. More recently  \citet{VC05} found a smaller 
deficit, $[M/H]=-0.7$,  which at any rate places it at least among  old disk 
stars. According to \citet{VC05}, the $\alpha$--elements are enhanced by a factor
$\sim$2 with respect to solar scaled abundances.
 
An excellent summary and a critical analysis of the dynamical data, collected 
over more than a century, has been recently published by \citet{Griff04} 
together with a long term radial velocity study, therefore we  report here only 
the strictly necessary information and refer to the quoted paper for a more 
complete review. The main updated orbital   parameters from various sources 
are collected in Table~\ref{orbit}. 
 
\begin{table} 
 \centering 
  \caption{Orbital and astrometric parameters of 85~Peg} 
  \label{orbit} 
  \begin{tabular}{@{}lllc@{}} 
  \hline  
parameter&	value & error	& ref. \\   
  \hline  
$P$ (yr) 			& 26.31  		&$\pm 0.01 $	& 2	\\ 
$a$					& 0\farcs 83	&$\pm 0\farcs 01$ & 1\\ 
$i$					& -49\degr		&$\pm 1\degr$	& 1	\\ 
$e$					& 0.38			&$\pm 0.01$		& 1	\\ 
$\pi$ (mas)			& 82.5 			&$ \pm 0.8$		& 1 \\  
\hline 
$K_1$ (km s$^{-1}$)	& 4.49 			&$\pm 0.05$		& 2	\\ 
$\gamma ($km s$^{-1}$)& -36.22 		& $\pm 0.03$	& 2	\\ 
$a_1 \sin i$ (AU)	& 3.68 			&$\pm 0.04$		& 2	\\ 
$f(\mathcal{M})$ (\massun)	& 0.0722 		&$\pm 0.024$	& 2	\\ 
 
\hline 
\end{tabular} 
 
\medskip 
Reference coding: (1) \citet{Soder99}, (2) \citet{Griff04}  
\end{table} 
           
\begin{table} 
 \centering 
  \caption{Physical parameters of 85~Peg} 
  \label{phys} 
  \begin{tabular}{@{}lllc@{}} 
  \hline  
parameter&	primary & secondary	& ref. \\   
  \hline  
\mass/\massun	& 0.77  $\pm 0.05$	& 0.72  $\pm 0.05$	& 1	\\ 
\teff (K)	& 5600  $\pm 50$	& 4200  $\pm 200$	& 2,3\\ 
\Mv  			& 5.39  $\pm 0.04$	& 8.47  $\pm 0.29$	& 4	\\ 
$L$ /\Lsun 		& 0.617 $\pm 0.02$	& 0.072 $\pm 0.03$	& 4	\\ 
$[M/H]$			& -0.7	$\pm 0.1 $	&					& 2 \\ 
$\log$ g		& 4.6	$\pm 0.1$	&					& 2 \\ 
\hline 
\end{tabular} 
\medskip 
 
Errors are standard deviations. 
Reference coding: (1) \citet{Griff04}, (2) \citet{VC05}, (3) \citet{fernandes}   
(4) present paper 
\end{table}

The often quoted \citep{Lipp81,Griff04} remark by \citet{Str55} on 85~Peg: 
``this double star is a thorn in the side of the present evolutionary theories 
of single stars and of widely separated double stars" essentially refers  to 
the discrepancy between the reduced mass 
($B=\mathcal{M}_2/(\mathcal{M}_1+\mathcal{M}_2)$), as derived from 
astrometry, and the value expected on the basis of the magnitude difference. 
The former is typically around 1/2, e.g. the recent value $B=0.528 \pm 0.034$ 
from Hipparcos results \citep{MM98}, and implies a secondary star equally or 
slightly more massive than the primary, the latter is around $B=0.38$. With  a 
well constrained total mass of \mass=$1.49 \pm 0.1$~\massun the secondary mass varies, 
then, from $0.77$ to $0.55$~\massun. 
 
The value of the reduced mass from astrometry is confirmed by the spectroscopic 
orbit of \citet{Griff04}, as his value of the mass function 
$f(\mathcal{M})=0.0722$~\massun,
when coupled to the total mass and inclination from visual 
observations,  provides  $B=0.482 \pm 0.019$, in good agreement with the 
abovementioned value.   A possible and straightforward explanation  is that the 
secondary is in its  turn a binary system, with an undetected  secondary 
(85~Peg~Bb) storing the  ``excess" mass (a few tenths of solar mass). This 
hypothesis was already put  forward by \citet{Hall48} and reappears in a number 
of subsequent works.  
  
The atmospheric parameters of 85~Peg have been as well the subject of several 
studies. An accurate determination of the primary effective temperature and 
chemical composition, by fitting of the Balmer line wings, was obtained by 
\citet{VV00}, as reported in \citet[][ hereafter FML02]{fernandes},  and  has 
recently been updated \citep{VC05}.  According to the latter work, the primary 
effective temperature  is \teff= 5600 $\pm 50$~K. 
Other, somewhat different, determinations   yield  also 
lower values;  \citet{Ful00} obtained the lowest one (\teff= 5275~K), but the 
discrepancy is  presumably due to the derivation of the temperature from the 
FeI lines under  the assumption of LTE.  \citet{TI99} showed that non-LTE 
effects can affect the  result in the case of sub-dwarfs and the same 
\citet{Ful00} finds indeed a  systematic effect, his temperatures being lower 
than those from other methods. 
    
The individual absolute magnitudes and relative errors were determined by us 
using the apparent magnitudes and relative errors given by \citet{tenBr00} who 
got $BVRI$ differential photometry of the components by means of adaptive 
optics, and derived the individual magnitudes from these and from the composite 
magnitudes found in \citet{MMH97}. We used, moreover, the Hipparcos parallax 
and errors as corrected by \citet{Soder99} for binarity effects. The bolometric 
corrections were obtained from the best fit coefficients of \citet{Flor96}. The 
physical parameters of the binary are collected in Table~\ref{phys}. Errors on 
luminosities do not take into account the uncertainty on bolometric correction 
which should be negligible at the primary temperature but could be relevant for 
the secondary component. 
 
FML02 determined by a least square algorithm the set of   physical parameters, 
with their evolutionary stellar code, that best fit the  available photometric, 
astrometric and spectroscopic data (under the hypothesis  of equal age and 
chemical composition of the components). They used, however, as input  to the 
procedure, a rather lower value for the reduced mass ($B=0.44$ vs. 
$B=0.482$ from Griffin 2004). Consequently  they determine a value for the 
primary mass, 0.88 \massun, which is  significantly larger than Griffin's one. 
These authors derive  a best fit age of 9.3 $\pm 0.5$~Gyr and a (today's) 
metallicity value  $\mathrm{[Fe/H]}=-0.55$. 

 
   
\section{Modeling} \label{models} 
 
We use the Aton 3.0 version of the Aton code (Mazzitelli 1979) whose main 
physical inputs are described in \citet{ventura-dantona}.  We describe here the 
properties of this version of the code which are  relevant to the  computation 
of the oscillation frequencies: namely the choice of opacities,  the modeling 
of convection and diffusion, and  the computation of an updated  equation of 
state. Only models having gray boundary conditions are used in this work.
 
\subsection{Opacities} 
We adopt the \citet{alexander} opacities at temperatures smaller than 12000~K, 
and the 
OPAL opacities, in the version documented by \citet{iglesias1996}. Opacities 
and  thermodynamics are computed in the code according to the prescriptions 
given in  the following Section.  
 
\subsection{Equation of state} 
 
Tables of equation of state (EOS) are computed outside the main code. For each 
metal abundance, represented by the mass fraction Z, six tables are built, with 
hydrogen content X from 0 to $1-$Z. The input variables are temperature T and 
gas pressure P$_{\rm gas}$, and the output quantities are the density $\rho$, 
specific heat at constant pressure C$_{\rm p}$, the ratio $\gamma=C_p/C_v$, the 
adiabatic gradient $\nabla_{ad}$, $\chi_\rho$\ and the exponents $\Gamma_1$, 
$\Gamma_2$\ and $\Gamma_3$. Simple functional relations then provide any other 
thermodynamic quantity. Tables are rectangular in the plane T--P$_{\rm gas}$, 
in order to simplify interpolation by means of third degree bidimensional 
spline functions. Notice that the regions of very high temperature and very low 
pressure are never used, and are filled up with simple formulations of the EOS, 
only to provide more convenient interpolation. The tables are filled up in 
three steps: first the thermodynamic quantities are computed according to the 
formulation by Stolzmann \& Bl\"ocker (1996, 2000) for fully 
ionized gas. This EOS allows to consider explicitly different elements 
mixtured. It includes degeneracy (both classic and relativistic), coulombian 
effects, exchange interactions and other corrections, and it is the most modern 
description available for ionized gas. It holds, however, only for high 
densities. 
  
The region in which pressure ionization is important is overwritten with 
\citet{saumon1995} EOS. In this way we fill the whole region  in which the gas 
is not ionized. However, \citet{saumon1995} EOS is only given for pure hydrogen 
and pure helium, without metals. Therefore the metallicity Z is ``simulated" 
through Y and the different hydrogen -- helium mixtures are interpolated 
through the additive volume law. This is particularly unsafe where strong 
configurational effects are expected, and the interpolation error may amount to 
$\sim 20$\% for normal H--rich mixtures \citep{montalban2000}. 
 
Use of \citet{saumon1995} EOS is limited to a small region of the T--P$_{\rm 
gas}$\ plane, by overwriting it with the OPAL 2001\footnote{the WEB version dated 
March 2002 is used} EOS \citep{rogers-eos1996,rogers01}, 
which is given for any Z  and ratio H/He. This EOS extends also to low pressures.  
 
After these large tables are written, for a given Z, six values of each physical 
quantities are computed for 6 different Y. A cubic unidimensional spline 
provides the interpolation for any input value of Y. The six tables for H/He 
and given Z are supplemented by fifteen tables He/C/O in which the EOS is 
directly  computed according to \citet{stolzmann1996} as the non ionized 
regions are not  present in stellar structure following helium ignition. The 
interpolation among  the fifteen tables is done by triangles in the plane C/O, 
as the stechiometric  condition is Y=1-X$_C$-X$_O$. 
 
In summary, the structure code computes the EOS quantities and the opacities 
memorizing the spline coefficients, and this allows to obtain full continuity 
both in the physical quantities and their first and second derivatives, a 
feature important when we deal with the computation of  multi-mode pulsations.  
  
\begin{figure} 
 \begin{center} 
	\includegraphics[width=84mm]{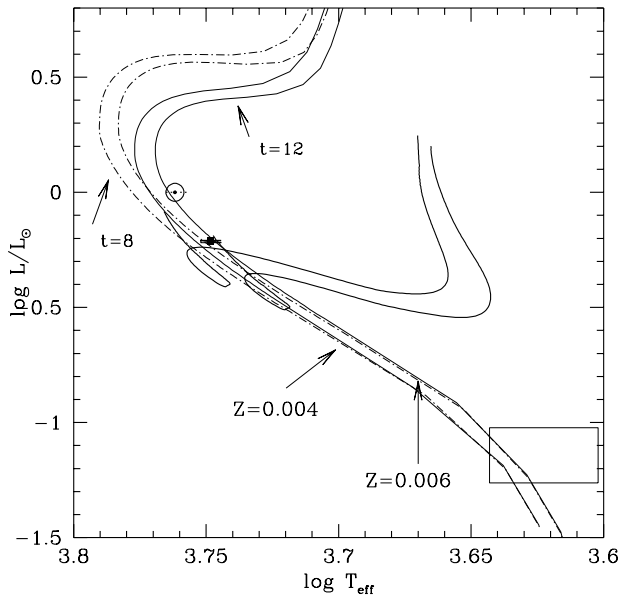} 
  \end{center} 
   
 \caption
{The location of 85~Peg~A (full dot with the small error box) 
 and 85~Peg~B (box) 
in the HR diagram, according to the  data in 
 Table~\ref{phys}. The present solar location is also shown. 
 We plot isochrones of $t=$8 (dash-dotted) and 12~Gyr 
 (dotted)  lines for metallicities  Z=0.004 and Z=0.006. For both sets of 
 models Y=0.24.  Two evolutionary tracks are also shown (solid lines), which bracket the 
 reasonable range of ages for the star: 0.791~\massun  (Z=0.004), which has an 
 age of about 8~Gyr at the luminosity of 85~Peg~A (model 4), 
and  0.775~\massun (Z=0.006), age $t \simeq 14$~Gyr (model 5).}  
\label{f1}  
\end{figure}

\subsection{Diffusion} 
 
The code includes  both gravitational and thermal helium  diffusion according 
to \citet{iben-mc1985}. Helium diffusion is a key ingredient in the  solar 
model, necessary to adequately reproduce the run of the sound speed as  derived 
from the extremely precise helioseismological data \citep{basu}.

\subsection{Convection} 
Convection in the code can be treated either according to the Full Spectrum 
Turbulence (FST) model by \citet{cm91} and \citet{cgm96}, or according to the 
classic mixing length theory (MLT) model  (B\"ohm-Vitense 1958). 
A detailed description of the convective model is given in \citet{ventura98}.
The FST model is characterized by convective fluxes in which the full eddies
distribution is accounted for, and by a convective scale length defined as the
harmonic mean between the distance of the layer from the top and the bottom of the
convective zone. In addition, the distance to the top (and bottom) is increased by a 
fraction $\beta H_p$, where $\beta$\ is calibrated in order to obtain the observed
solar radius at the solar age (in the range $\beta=0.1 -0.2$). Non 
instantaneous mixing is generally used  as described by \citet{ventura98}. 
The mixing is treated as a diffusion, with the diffusion velocity obtained
in the framework of the FST model. This mixing  is very relevant for hydrogen 
or helium burning in convective cores, but it is not strictly important in this work, 
as we will be dealing only with burning of  hydrogen in 
radiative cores. 
 
\subsection{Numerical resolution} 
Particularly to compute the solar structure to be compared with the very 
precise results of helioseismology, we take care to have a very good numerical 
resolution of the structure, both in the temporal evolution (about 1000 time-
steps are taken to reach the solar age) and spatial resolution (about 2000 mesh 
points are adopted). For the exploratory models relative to 85~Peg~A presented 
in  this work, we only employ $\sim$800 mesh points and broader temporal 
evolution. In any case we took care of having an appropriate distribution of 
meshes both in the central region (to have a careful computation of nuclear 
evolution), in the overadiabatic region and close to the bottom of the external 
convective region (to have a careful computation of the oscillation 
frequencies).
  
\subsection{Use of solar-scaled abundances} 
In the code, we did not yet implement the use of $\alpha$--enhanced opacities 
and EOS. At present, the uncertainties both in the metallicity determination 
and in the metal diffusion (see next section) are such that this improvement is 
not yet necessary, although it must be planned for the future. For population 
II stars, we follow the suggestion by Salaris et al. (1993), who have shown 
that the effect of the increase of $\alpha$--elements abundances in the energy 
production and opacities can be mimicked by increasing the total metal 
abundance. 
 
\section{Metal diffusion in present evolutionary codes} \label{metdif} 
 
The current version of the evolutionary code does not include metal diffusion, 
whose  effect is however much controversial. In fact, the same {\sl ``sign''} 
of metal  diffusion is not yet clear, particularly in the context of low 
metallicity  stars. The problem is widely discussed by \citet{gratton-sneden}. 
First of all,  the theoretical results differ according to the description 
employed: most  results based on the description by  Thoul, Bahcall, \& Loeb 
(1994), which is widely used, predict for the metallicity of the cluster NGC 
6397 a depletion in [Fe/H] by 0.2 -- 0.4~dex (depending on the age) at the 
turnoff. On the other hand, the  recent models by \citet{richard}, that take 
into account the effect of partial  ionization and radiative accelerations, 
show that whereas some elements like He and  Li are expected to be depleted, 
others (like Fe) are expected to be  significantly enhanced. From the 
observational point of view,  \citet{gratton2001} find that turnoff and 
subgiant stars in the globular cluster NGC 6397  have the same metal abundance 
within 0.1~dex. This is neither consistent with models which predict a strong 
diffusion in main sequence, nor with \citet{richard} models, as there should be 
a large difference between the turnoff abundance --where the metals have been 
either partially depleted from the whole convective region, or enhanced-- and 
the subgiants, where deep convection should have restored the initial 
metallicity.  
 
In any case, standard models predict that the effect of diffusion at the 
turnoff of globular cluster stars increases when decreasing the metallicity, as 
the main sequence shifts to increasingly hotter temperatures, for which the 
external convective layers are smaller, providing higher diffusion velocities 
at their bottom. Therefore, a negligible metal diffusion is expected for 
85~Peg~A, in view of the very small --if any-- effect present in the much more 
metal poor cluster NGC 6397. 
 
In view of these results, it is difficult to accept such a large effect as 
produced in the models by FML02, which start with a quasi--solar metallicity 
($\mathrm{[Fe/H]}= -0.185$) to arrive at the present metallicity, assumed to be 
$\mathrm{[Fe/H]}= -0.55$. In particular, these models would require that the 
high velocity giants in Hipparcos catalogue have on average metallicities 
larger by a factor two than their turnoff or main sequence counterparts, a 
feature which has not been observed. 

Nevertheless, in order to quantify the effect of the metal diffusion, we have 
also considered models computed by employing the Code Li\'egeois d'\'Evolution 
Stellaire (CLES) in which the helium and metal diffusion has been implemented 
according to \citet{thoul}. Comparisons between models calculated with and 
without metal diffusion will be shown in Sec. 7.2. 
 
 \begin{table*} 
 \centering 
  \caption{Selected models for 85~Peg~A} 
\label{model}  
  \begin{tabular}{clccccccc} 
  \hline 
 mod. &convection & \mass/\massun &    Z    &    t(Gyr) & L/\Lsun & R/R$_\odot$ &\teff& 
   X$_c$  \\ 
 \hline 
 1 &FST&  0.750    & 0.004 & 13.80 & 0.616 & 0.817 &5661  & 0.1670 \\ 
 2 &FST&  0.763    & 0.004 & 11.96 & 0.617 & 0.810 &5689  & 0.2103 \\ 
 3 &FST&  0.778    & 0.004 &  9.97 & 0.619 & 0.803 &5716   & 0.2669 \\ 
 4 &FST&  0.791    & 0.004 &  8.31 & 0.618 & 0.797 &5737   & 0.3240 \\ 
 5 &FST&  0.775    & 0.006 & 13.97 & 0.619 & 0.846 &5572  & 0.1653 \\ 
 6 &FST&  0.788    & 0.006 & 12.10 & 0.618 & 0.841 &5587  & 0.2100 \\ 
 7 &FST&  0.803    & 0.006 & 10.08 & 0.618 & 0.833 &5615  & 0.2675 \\ 
 8 &FST&  0.820    & 0.006 &  7.88 & 0.616 & 0.823 &5644   & 0.3443 \\ 
 9 &MLT&  0.803    & 0.006 & 10.02 & 0.615 & 0.833 &5608  & 0.2706 \\ 
\hline 
\end{tabular} 
\end{table*}

\section{85~Peg in the evolutionary context} \label{85peg} 
 
We computed, by using the Aton code, two sets of evolutionary tracks 
for 85~Peg~A and B
with helium mass fraction  $Y= 0.24$ and metal mass fraction 
Z=0.004  and Z=0.006 corresponding, respectively, to solar scaled abundance  
$\mathrm{[Fe/H]}\sim-0.70$ \citep{VC05} and $\mathrm{[Fe/H]}\sim-0.55$ (FML02).
Considering these two different metallicities, we also get a first 
understanding of the uncertainty due to the adoption of solar scaled 
abundances in our code. The cross section are computed according to the NACRE
compilation \citep{angulo}.
All the models apart from model 9 in Table \ref{model} adopt the FST treatment of
the convective transport of energy in the interior, with a fine tuning
parameter $\beta= 0.17$.
In order to quantify the effect of the treatment of convection on the pulsational 
properties of the star, we also computed a track using the B\"ohm--Vitense
mixing length formalism (MLT) with $\alpha_{MLT} = l/H_p= 1.7$ (model 9). The
$\alpha$\ value has been chosen so that the MLT and FST tracks have the same location
in the HR diagram.

According to its space velocity, 85~Peg belongs to a group of moderate - high 
velocity stars, whose age is statistically limited to be larger than 
$\sim$8~Gyr  \citep{caloi}. We have then considered 8$\simlt$t$\simlt$14~Gyr as 
possible ages for 85~Peg.  In Table \ref{model} are collected, for each Z,  the 
mass  of the model  reaching the luminosity observed for 85~Peg~A 
(see Table~\ref{phys}) at 
the assumed ages of about 8, 10, 12 and 14~Gyr.  Fig. \ref{f1} shows the 
locations of 85~Peg~A and B according to the data  from Table~\ref{phys},  the 
8 and 12~Gyr isochrones for the different  metallicities and two extreme 
evolutionary tracks (model 4 and 5).  We notice that all the computed 
isochrones are compatible with the location of 85~Peg~B. The 12~Gyr isochrones 
of both metallicities are consistent with the \teff\ location of 85~Peg~A, 
while the isochrone of Z=0.004 and $t= 8$~Gyr has a too large \teff. However, 
we can consider as well this last isochrone on the basis of uncertainties on 
temperature.  On the observational side, the error of only $\pm 50$~K assigned 
to the \teff\ determination certainly does not include consideration of 
possible systematic errors. From the stellar structure modeling side, we do 
know that precise \teff's can not be obtained from first principles. In 
particular, we employ grey atmosphere models: \citet{montalban2001} have shown 
that non gray models for low metallicity predict slightly lower \teff\ than 
gray models (by $\sim 80$~K at Z=0.0002). Our results adopt the FST convection 
model, but a convective model with a slightly smaller efficiency may  provide 
smaller \teff. In summary, the models for Z=0.004 (the best value for the 
metallicity determined from spectra) and ages smaller than 12~Gyr  have also 
full right to be considered in relation to 85~Peg, as we will check by looking 
at the mass  luminosity relation. 

\begin{figure} 
 \begin{center} 
    \includegraphics[width=84mm]{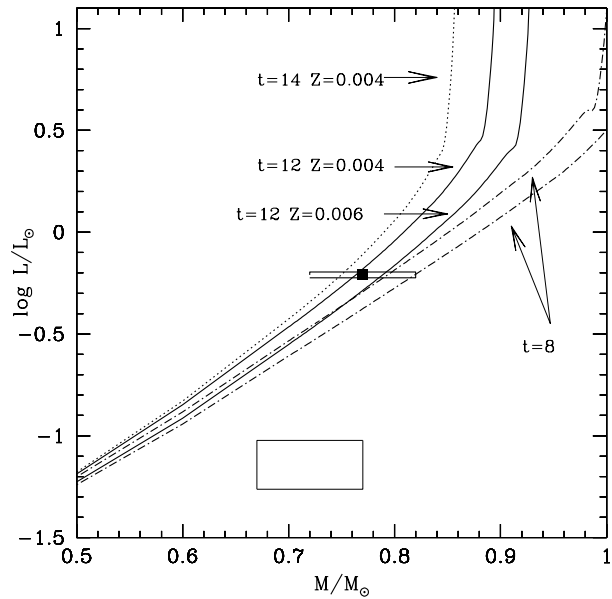} 
 \end{center} 
 
 \caption{The masses and luminosities of 85~Peg~A and 85~Peg~B, 
as derived by \citet{Griff04}, are compared 
 with the mass luminosity relation of the computed isochrones shown in 
 Fig. \ref{f1} and with that for t=14~Gyr and Z=0.004 [model 1].  
 If one excludes ages larger than $\sim 14$~Gyr, the minimum acceptable mass 
 is $\sim 0.75$~\massun.} 
\label{f2} 
\end{figure} 
 
The mass luminosity relation provides a better information than the  HR 
diagram, as it is not affected by the uncertainty in \teff. The mass luminosity 
relations of the computed isochrones shown in Fig. \ref{f1} are given in Fig. 
\ref{f2} with also that for $t \simeq 14$~Gyr and Z=0.004.  The location of 
85~Peg~A according to the data in Table~\ref{phys},  taking into account the 
uncertainties, is in good agreement with all the computed Z=0.004 or Z=0.006 
isochrones. Besides, in the hypothesis of an age smaller than 14~Gyr, we can 
put a lower limit to the primary mass at 0.75 \massun. In summary, we can 
consider for our analysis all the models at the luminosity  of 85~Peg~A, from 
0.75~\massun (Z=0.004), which has an age of $\sim 14$~Gyr to  0.82~\massun 
(Z=0.006) which has an age of $\sim 8$~Gyr. 

Notice the location of  85~Peg~B in 
the figure: indeed its dynamical mass is much larger than any  acceptable value 
at that given luminosity. If we consider the value given by \citet{Griff04}, 
the dynamical mass of the secondary component is $\sim 0.2$~\massun\ larger 
than the  possible main sequence mass. 
In the hypothesis that 85~Peg~B is in its turn a binary system, its primary
component, 85~Peg~Ba, should have a mass of $\approx$ 0.52 \massun (depending 
on the  age), and its companion, 85~Peg~Bb, of about 0.2  \massun. Accordingly, the
luminosity of component Bb would be $\approx$ 0.0085 \Lsun and that of
the primary of 0.064 \Lsun (as the value in Table \ref{phys} would refer to
the composite luminosity).
The binary nature of 85~Peg~B could be checked by spectroscopic or 
interferometric observations (depending on its unknown period), if an
instrument capable of separating the components of the visual binary
is used, such as an adaptive optics telescope. The various possibilities in relation 
to the expected orbital period have already been examined in \citet{Griff04}.

In the next Sections the models are taken as input to derive  
the oscillation spectrum.  
 
\section{Pulsation analysis} \label{puls}   
 
We use the adiabatic oscillation code by J{\o}rgen Christensen-Dalsgaard 
available in the WEB to calculate the p-mode eigenfrequencies. Some analytical 
and numerical aspects of oscillation codes are described in \citet{CD91}.   
 
From the asteroseismic point of view 85~Peg~A can be classified as a solar-type 
star,  a main-sequence star in which oscillations are excited stochastically by 
vigorous near-surface convection in a broad spectrum of low amplitude p-modes, 
as in the Sun.  
The interpretation of the pulsational properties of a solar-type star  can be  
obtained by considering the asymptotic theory \citep{Tassoul80}. This predicts  
that the oscillation frequencies $\nu_{n l}$ of acoustic modes, characterized by  
radial order $n$ at harmonic degree $l$,  should satisfy the following  
approximation:    
\begin{equation}    
\nu_{n l}=\Delta\left(n+\frac{l}{2}+\alpha+\frac{1}{4} \right)    
+\epsilon_{n l} \; ,    
\label{eq1}    
\end{equation}    
where $\alpha$ is a function of the frequency determined by the properties of  
the surface layers and $\epsilon_{n l}$   is a small correction which depends on  
the conditions in the stellar core. $\Delta$ is the inverse of    the sound  
travel time across the stellar diameter:  
\begin{equation} 
\Delta={\left(2\int_{0}^{R}\!\frac{{\rm d}r}{c}\right)}^{-1},  
\end{equation} 
where $c$ is the local speed of sound at radius $r$ and $R$ is the photospheric  
stellar radius. The property expressed by Eq.~(\ref{eq1}) may provide almost  
immediate asteroseismic inferences on stellar parameters and constraints on  
theoretical models for a variety of solar-like stars in a wide range of  
evolutionary stages.   For a given $l$, the acoustic spectra show a series of  
equally spaced peaks between p modes of same degree and adjacent $n$, whose  
frequency separation represents the so called large separation which is  
approximately equivalent to $\Delta$: 
\begin{equation} 
\Delta \simeq\nu_{n+1 \: l}-\nu_{n l}\equiv\Delta_{l}\,. 
\label{EQ_3} 
\end{equation} 
 The spectra are characterized by another series of peaks, whose narrow  
 separation is $d_{l \: l+2}$, known as small separation: 
\begin{equation} 
d_{l \: l+2}\equiv \nu_{n l}-\nu_{n-1 \: l+2}=(4l+6){\rm D}_{0} 
\label{EQ_4} 
\end{equation} 
where 
\begin{equation} 
{\rm D}_{0}=-\frac{\Delta}{4\pi^{2}\nu_{n l}}\left[\frac{c(R)}{R}- 
\int_{0}^{R} 
\frac{{\rm d}c}{{\rm d}r}\frac{{\rm d}r}{r}\right] \, . 
\end{equation} 
$\Delta$, and hence the general spectrum of acoustic modes,  scales 
approximately as the square root of the mean density, that is,   $\Delta 
\propto \sqrt{\hat{\rho}}= \sqrt{\mathcal{M}/R^{3}}$. On the other hand, the 
small frequency separations are  sensitive to the chemical composition gradient 
in   central regions of the  star and, hence, to its evolutionary state. Thus, 
the determination of both  large and small frequency separation, $\Delta_l$ and 
$d_{l \: l+2}$,  provides measures of the mass and of the age of the star 
\citep [e. g.][]{CD88}.    

 \subsection{Sub-surface effects and the ratio of small to large separation} 
\label{subsur}
 
The seismic analysis of observed acoustic frequencies is an extremely powerful tool  
for the investigation of the internal structure of the stars, but the use of  
large and small separations can be misleading if not accurately considered.  
Theoretical pulsation frequencies, essential for asteroseismic investigation,  
are calculated on theoretical models of stars which are inevitably affected by  
errors. In particular, the structure of the near-surface regions of stars,
which strongly affects the large separation, is  
quite uncertain. In fact, there are still substantial ambiguities in 
the theoretical description of: \,  
i) the convective flux, \, ii) the mechanisms of excitation and  
damping of the oscillations, \, iii) the equation of state to  
describe the thermodynamic properties of the stellar structure,   
and \, iv) the non-adiabatic effects. 
Limiting the investigation to the use  
of small separation does not solve the problem, since the small separation,  
which  is determined principally by the conditions in the core, also retains some  
sensitivity to the mean density and to the detailed properties of the stellar  
envelope. 
 
To solve these problems, the use of a new seismic indicator was introduced by  
\citet{RV03}, who compared the pulsating properties of several  
models with exactly the same interior structure, but with different outer envelopes.  
They showed that the ratios of small to large separations,  
\begin{equation} r_{l}=d_{l \: l+2}/\Delta_{l}\, , \end{equation} are  
independent of the structure of the outer layers and, hence, can be used as  
diagnostic of the interior of stars. 

\subsection{Echelle diagram}

In order to study the large and small separations of the observed and computed
oscillation frequency spectra,  it is common to use the {\it echelle diagram}
\citep{Grec83}.   In this diagram each frequency is expressed in terms of  an
integer multiple of $\Delta$, according to

\begin{equation}
\nu_{n l}=\nu_0+k\Delta+\tilde{\nu}_{n l} \; ,
\end{equation}
where $\nu_0$ is an arbitrary reference frequency, $k$ is an integer and
$\tilde{\nu}_{n l}$ is a residual frequency which   lies in the range $0-
\Delta$.   Plotting $\nu_0+k\Delta$ against
$\tilde{\nu}_{n l}$ one expects, according to the asymptotic 
theory (Eq. \ref{eq1}), that computed frequencies
fall in parallel columns, one for each
harmonic degree. The distance between consecutive frequencies of same harmonic
degree, belonging to the same column, represents the large separation; modes of
even degree (0 and 2) and modes of odd degree (1 and 3) fall in two pairs of
closely spaced columns, the distance between the columns in each pair
representing the small separation.

\begin{figure}
 \begin{center}
        \includegraphics[width=84mm]{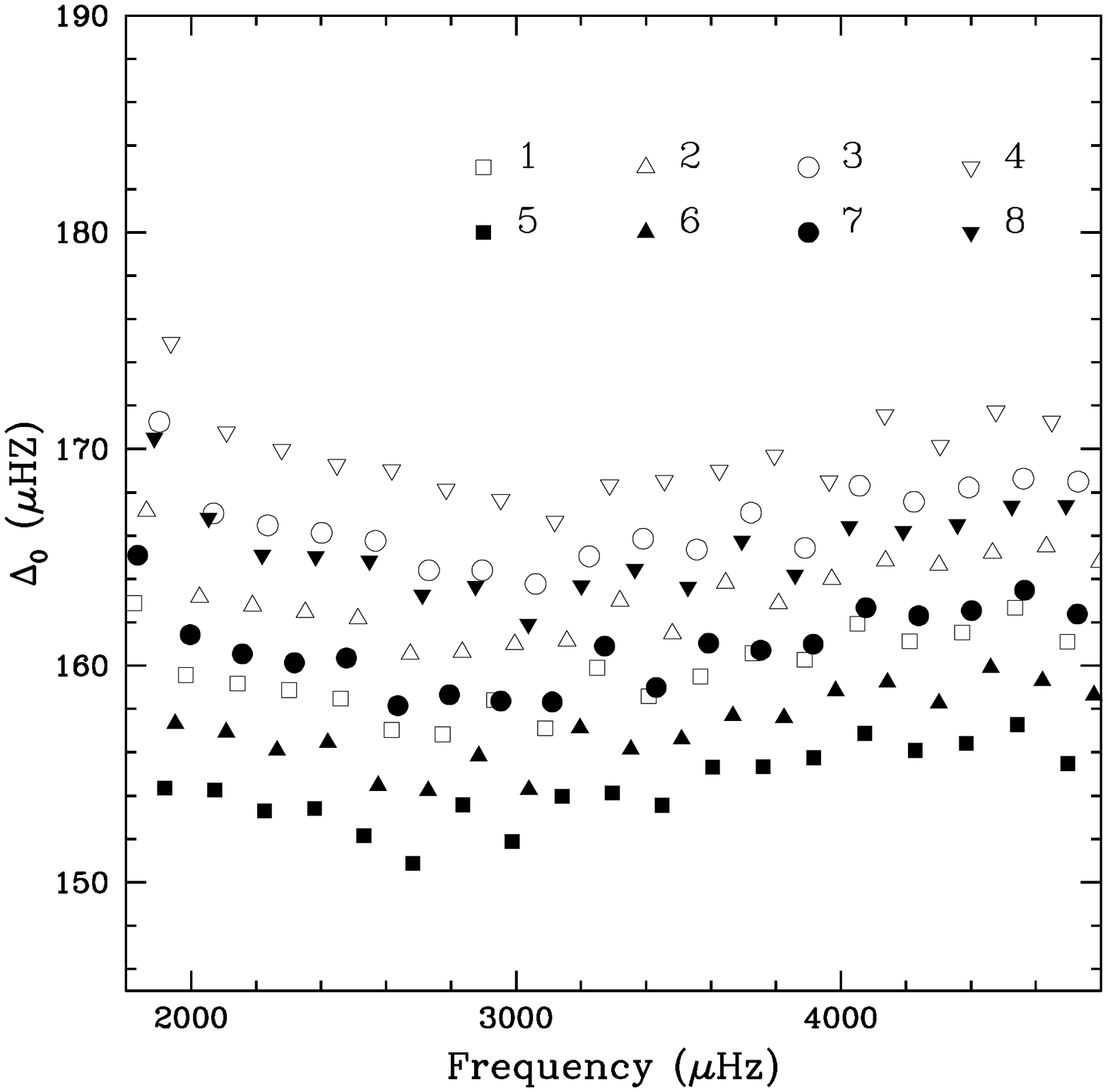}
        \includegraphics[width=84mm]{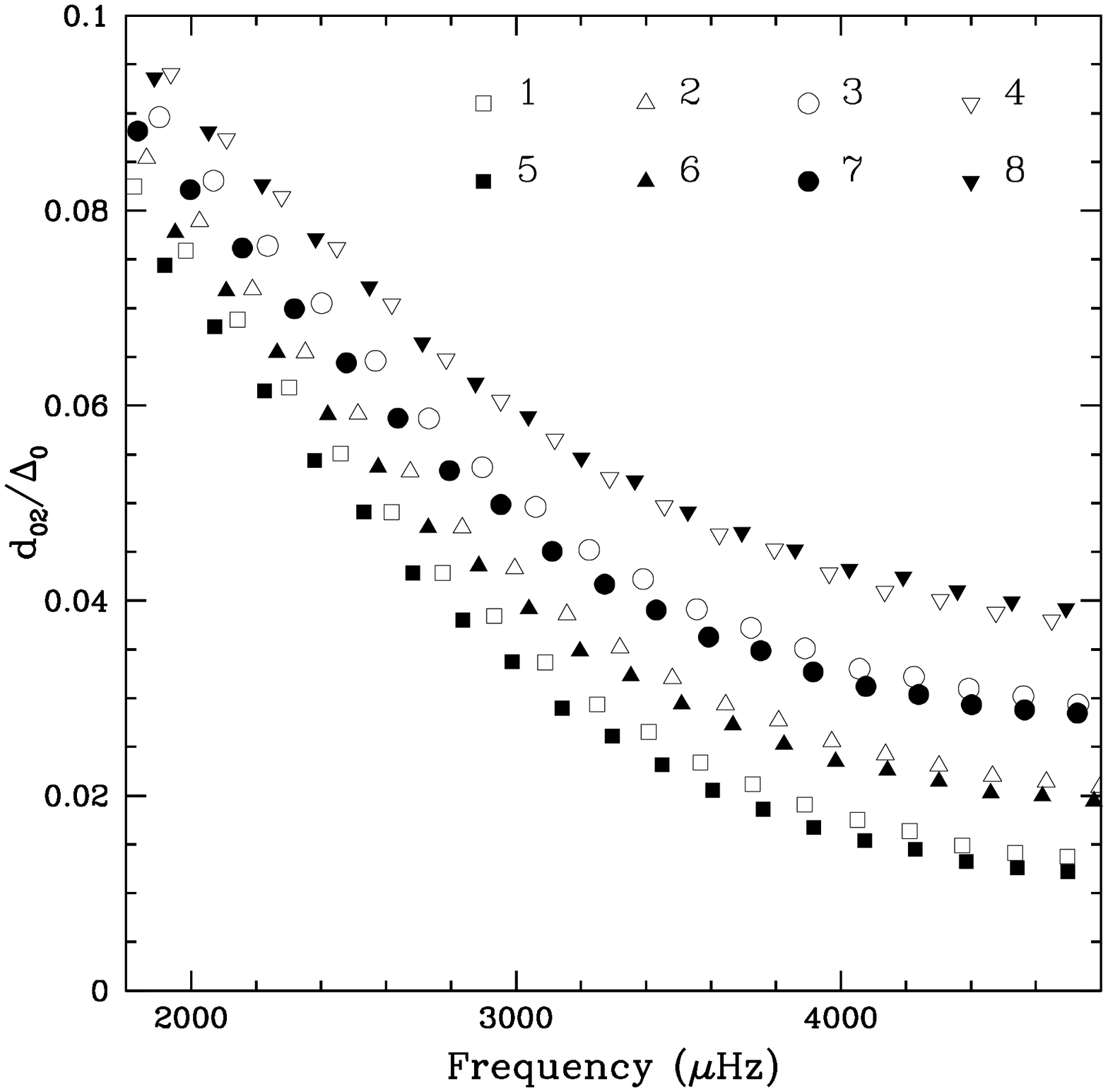}
      \end{center}
 \caption{
Open symbols refer to models with Z=0.004, filled symbols to Z=0.006, 
equal symbols correspond to models of similar age.
Upper figure: large separation $\Delta_0$ for $l=0$
as function of the frequency calculated for selected models described 
in Table~\ref{model}.
Lower panel: ratio of small $d_{0 2}$ to large separations $\Delta_0$ between 
$l=0$ and $l=2$ for eight models of Table~\ref{model}.}
 \label{f3}
\end{figure}


\section{Prediction of the oscillation frequencies of 85~Peg~A} \label{oscill}

We computed eigenfrequencies for all the selected structure models described in 
Table~\ref{model} and consistent with the observed basic parameters of the 
primary component of 85~Peg. Since the oscillation modes which can be detected 
in stars are generally characterized by low harmonic degree (owing to the 
point-like character of the sources), we limited our analysis to the 
calculation of  modes with harmonic degrees $l=0,1,2,3$. Theoretical 
calculations on the selected models predict that 85~Peg~A is pervaded by 
acoustic modes with frequencies in the range $1000-5000\,\mu$Hz for harmonic 
degrees $l=0-3$. The oscillation spectrum is characterized by a large 
separation for the radial modes $\Delta_0$ of about $150-170  \, \mu\mathrm{Hz}$ 
(Fig.~\ref{f3}).

\begin{figure}
 \begin{center}
        \includegraphics[width=84mm]{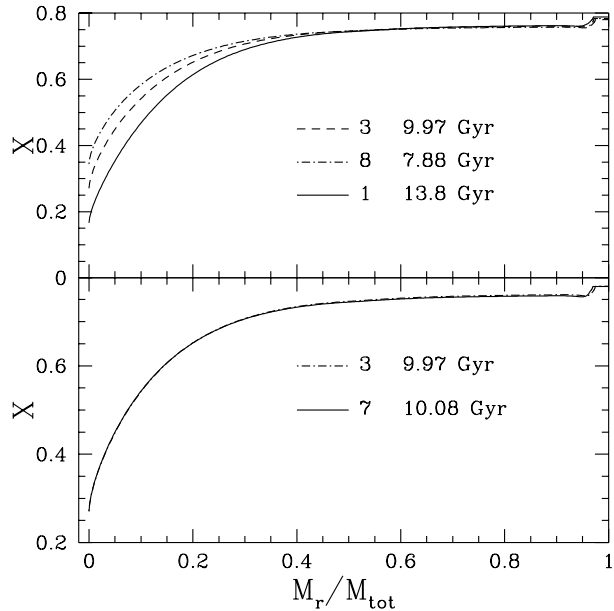}
  \end{center}
 \caption{The top figure shows the hydrogen mass fraction X along the 
 structure of selected models for increasing age. The bottom figure shows that
 the two models having age close to 10~Gyr have an almost equal hydrogen
 profile in spite of the different metallicity.}
\label{f6}
\end{figure}

As predicted by theory, the value of the large separation decreases as the mean 
density decreases. For fixed metallicity, the large separation increases with 
the mass. In fact, the large separation strongly depends on the condition on 
the surface, and hence on the value of the metallicity. This means that the 
value of the mass cannot be estimated from the observed large separation alone.

It is very instructive to
 consider the behaviour of the large separation as a function of frequency (Fig.
~\ref{f3}).
Consistently with the solar case,
the theoretical $\Delta_l$
shows an evident oscillatory dependence on frequency.
 Such quasi-periodic behaviour of the
frequencies rises from discontinuity's regions
 localized at certain acoustic depth in the structure.
This peculiar property of the oscillation frequencies can be used
 to infer the helium abundance in the
 stellar envelope or to determine the properties and the location
 of the base of
the convective envelope \citep[e.g][]{Ba04, maz01, mont98, mon94, mon98, mon00,
mon02}.

Fig.~\ref{f3} employs the parameter $r_0=d_{02}/\Delta_0$ devised by 
\citet{RV03}, in order to subtract  the contribution of the surface layers, 
yielding a diagnostic of the stellar interior alone. In fact, the difference in 
$d_{02}/\Delta_0$ among the models is easily understood in terms of the 
different internal structure: independently of the metallicity of the models, 
it is the core hydrogen content --and thus the age reached at the luminosity of 
85~Peg~A-- which determines $d_{02}/\Delta_0$. Thus, models in the same 
evolutive state have the same value of $r_l$. The plot of hydrogen mass 
fraction X inside some of the examined structures is shown in Fig. \ref{f6}.

It is clear that while measurements of small separation (or ratio $r_{l}$) will 
be useful to define the evolutive state of the star, the large separation will 
be employed to identify, among structures of same age, elements abundances and 
mass of the model which better reproduces the observations.

\subsection{MLT versus FST models}

\begin{figure} 
 \begin{center} 
	\includegraphics[width=84mm]{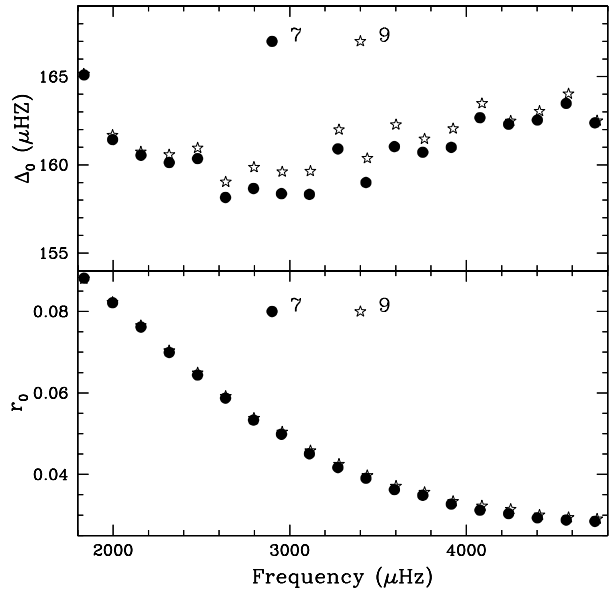} 
  \end{center} 
   
 \caption{Large separation $\Delta_0$ for $l=0$ (top figure) and ratio of small 
 to large separation (bottom)
 for model 7 and 9 of Table 3, having same mass and radius and nearly the same 
 age of 10~Gyr.
 The models differ in the modelling of the external convective layers: FST for model 7 and
 MLT for model 9.}
\label{mlt}
\end{figure}

While we adopt the FST model as a  standard for the overadiabatic convection, 
we have made a comparison between model 7, computed with FST, and model 9, 
computed with the MLT. We adopt $\alpha=l/H_p$=1.7, which provides a track 
having the same \teff\ location  of the FST track. The  comparison is made in 
Fig. \ref{mlt}: there is a difference of about $1 \mu$Hz in the large 
separations for $l=0$, while  the ratios $r_0$\ of the two models are very 
similar. Thus we conclude that the convection treatment does not affect the age 
determination by the seismic analysis. 

\subsection{The metal diffusion}

\begin{figure} 
 \begin{center} 
	\includegraphics[width=84mm]{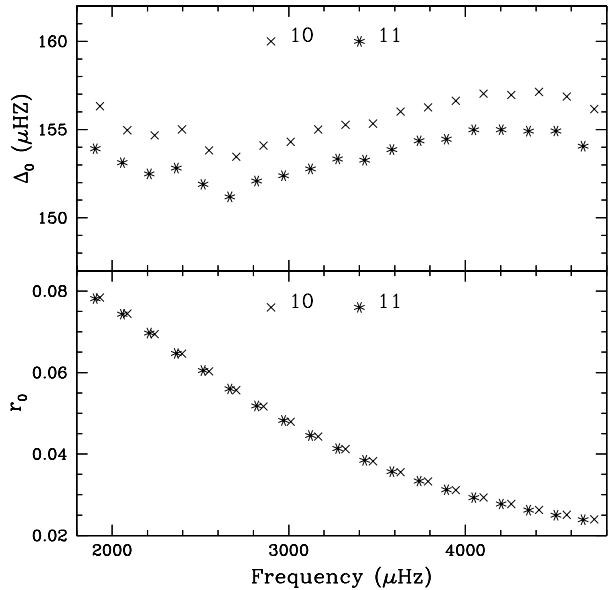}  
  \end{center} 
 \caption{Comparison between the CLES models showing the effect of metal diffusion
 on the large separation (top) and on the ratio of small to large separation (bottom). }
\label{josef}
\end{figure}

The ATON code does not include metal diffusion, so we adopt an entirely 
different code to have an idea of the influence of this physical input on the 
structure. The stellar structure and evolution code CLES adopts OPAL 2001 EOS; 
OPAL96 opacity tables plus Alexander \& Ferguson (1994) at T $<$ 6000 K; 
nuclear reaction rates from Caughlan \& Fowler (1988); MLT convection 
treatment; atmospheric boundary conditions given by Kurucz (1998) at T = \teff. 
Microscopic diffusion of all the elements is computed by using the subroutine 
by \citet{thoul}.  A detailed comparison of the results of the CLES and ATON 
codes is out of the purpose of this paper, although it is in our research 
projects. We list in Table 4 the physical parameters of two models in the error 
box of 85~Peg~A. We choose a  large age (12~Gyr) to make the comparison, in 
order to maximize the effect of diffusion. The initial metallicity is Z=0.006, 
corresponding to [Z/X]$_0$=--0.555. Model 11 includes only helium 
diffusion\footnote{Notice that the CLES model has X$_c$=0.20 in the center, 
while the corresponding ATON model 6 has X$_c$=0.21, in spite of being 
slightly older. 
The treatment of helium diffusion differs in the two codes. The description in 
the ATON code \citep{iben-mc1985} provides smaller diffusion velocities and 
this leads to this difference in X$_c$\  after 12~Gyr.}, while model 10 
includes both helium and metal diffusion. At the age of 12~Gyr [Z/X]$_s$, the
vale at the surface of 
model 10  has been reduced by $\sim$33\%\footnote{Notice that this decrease in 
[Z/X]$_s$\ in the surface layers is due both to the decrease in Z and to the 
increase of X due to the helium diffusion (this can be seen, e.g. in Fig. 5 
and from the change of [Z/X]$_s$\ also in model 11 which has constant Z).}, a 
relatively mild effect which confirms the prediction made in Section 4 that 
metal diffusion on a star having the \teff\ of 85~Peg~A should not be dramatic. 
 We show in Fig. \ref{josef} 
the oscillation properties of two models. We can see that the large separations 
differ by $\sim 2 \mu$Hz, but the ratio $r_0$\ is again 
very similar. Thus we conclude that the uncertainty in the computation of metal 
diffusion in the case of 85~Peg~A is totally contained into the uncertainty in 
metallicity, and does not affect the age determination.

 \begin{table*}
 \centering
  \caption{Selected models for 85~Peg~A obtained with CLES code
with and without metal diffusion}
\label{josefmod}
  \begin{tabular}{clccccccccc}
  \hline
 mod. &convection & metal diff.&\mass/\massun &   [ Z/X]$_{0}$    & [Z/X]$_{s}$ &    t(Gyr) & L/\Lsun & R/R$_\odot$ &\teff&
   X$_c$  \\
 \hline
 10 &MLT&YES&  0.788    & -0.555 &  -0.678 &   11.90 & 0.615 & 0.843 &5572  & 0.2035 \\ 
 11 &MLT&NO&  0.788    & -0.555 &   -0.583 &   11.96 & 0.611 & 0.851 &5538  & 0.2036 \\
\hline 
\end{tabular}
\end{table*} 

\subsection{Echelle diagrams}

\begin{figure}
 \begin{center}
 \includegraphics[width=84mm]{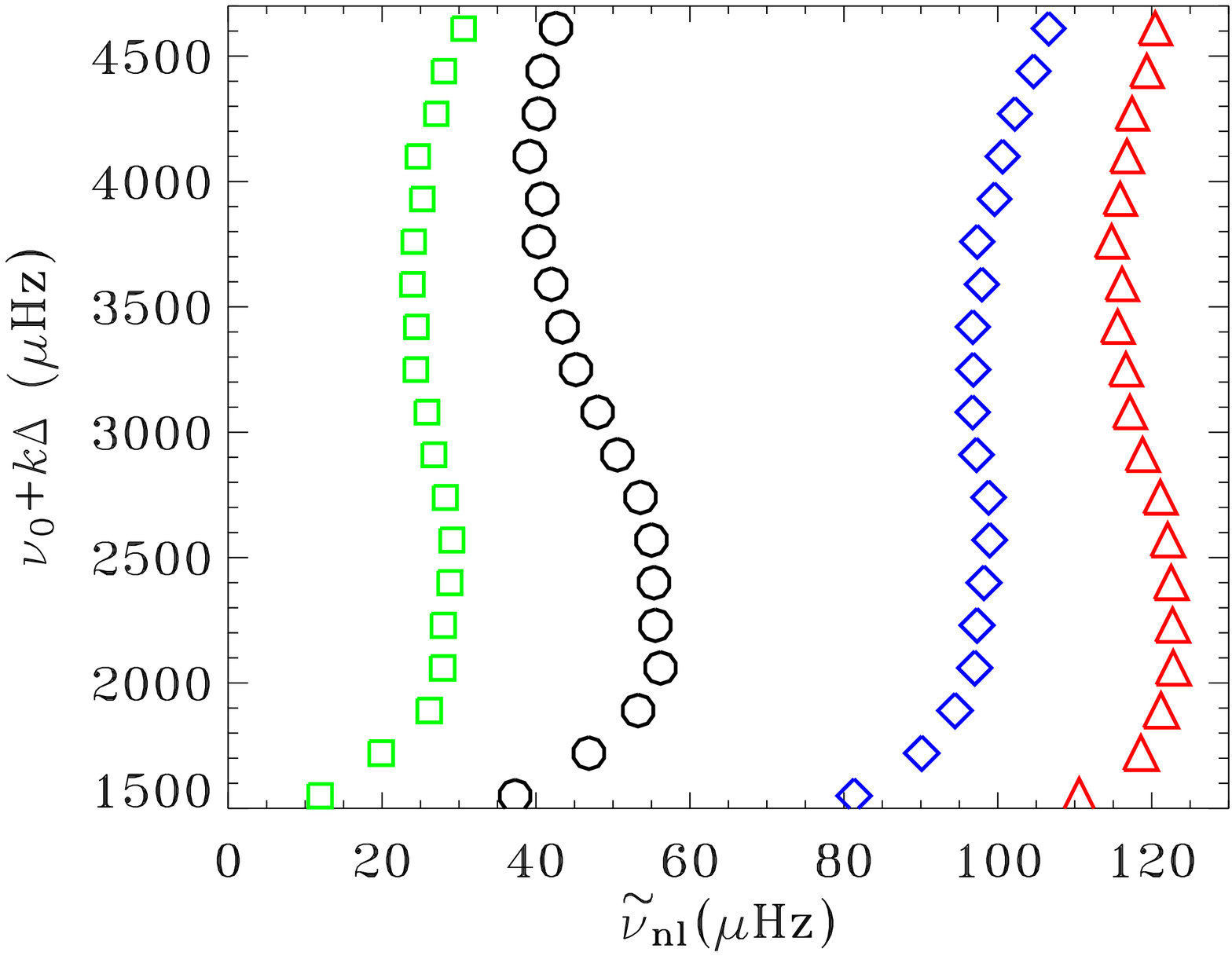}
 \includegraphics[width=84mm]{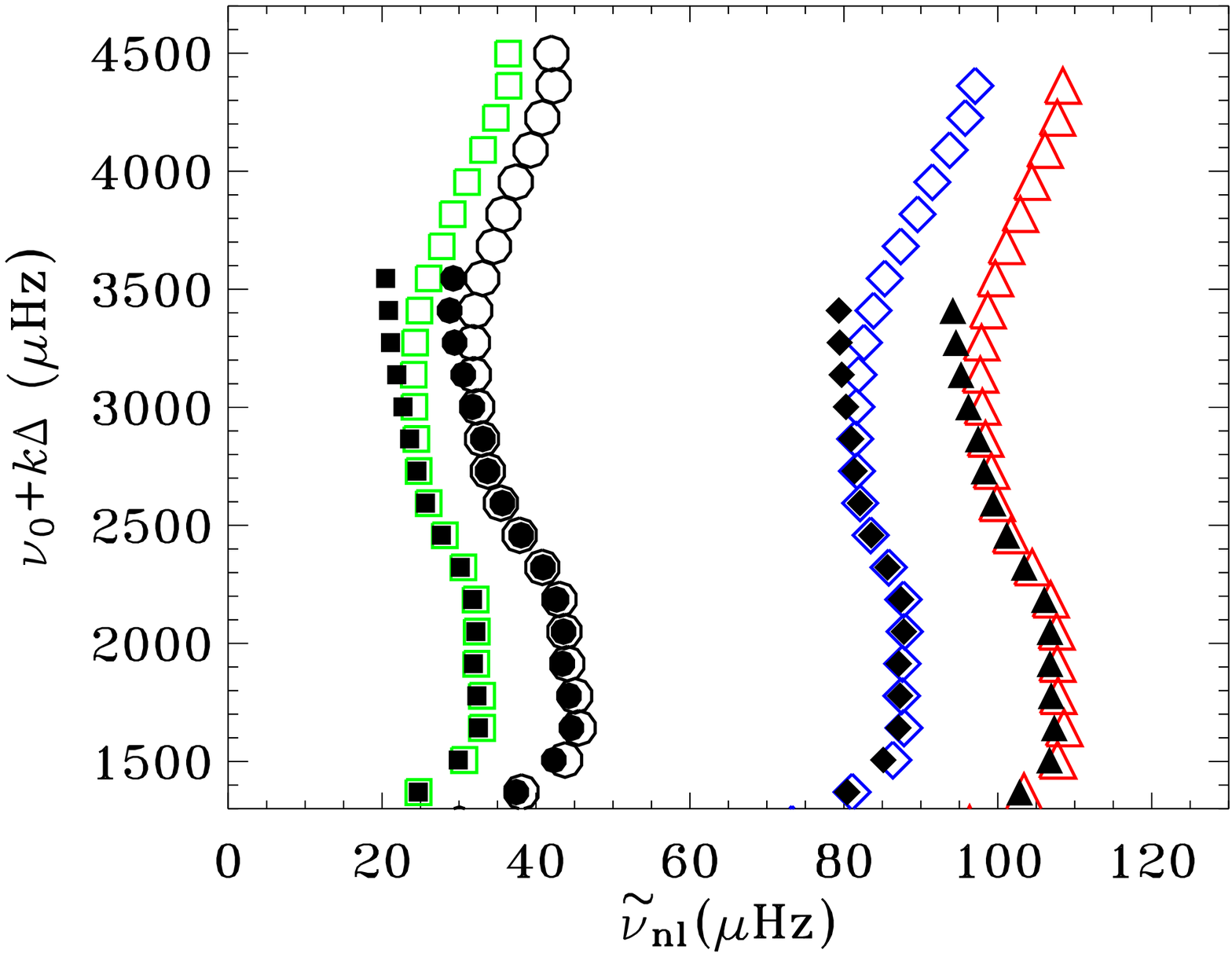}
  \end{center}
 \caption{Echelle diagrams respectively for
Model 4 (upper panel) of 85~Peg~A and a
model of the Sun
together with the observational results (lower panel).
The filled symbols show observed frequencies, while
the open symbols represent the computed frequencies.
Circles are used for modes with $l=0$, triangles for $l=1$, squares
for $l=2$, and diamonds for $l=3$.
Theoretical frequencies of 85~Peg~A are
plotted with $\Delta =170 \, \mu\mathrm{Hz}$,
 and $\nu_0=870\, \mu\mathrm{Hz}$.
  Theoretical and observed frequencies
of the Sun are
plotted with $\Delta =136 \, \mu\mathrm{Hz}$ and
 $\nu_0=690\, \mu\mathrm{Hz}$. }
\label{f7}
\end{figure}

In Fig. \ref{f7} we compare the echelle diagram for model~4 of 85~Peg~A
with that of a standard solar model having Y=0.28 and Z=0.02. The theoretical
echelle diagram of 85~Peg~A has been obtained by using the value of
$\Delta=170\, \mu\mathrm{Hz}$, and $\nu_0=870\, \mu\mathrm{Hz}$.
 The lower
panel  of Fig.~\ref{f7} obtained for the Sun, is drawn for both
the theoretical and the observed frequencies by using the value of
$\nu_0=690\, \mu\mathrm{Hz}$ as a reference frequency and a large separation
$\Delta =136\, \mu\mathrm{Hz}$ corresponding to the observed average which
can be calculated on MDI/SOHO frequencies (Schou 1998).
Like in the Sun, the theoretical
echelle diagram of 85~Peg~A shows a regular behaviour of
the frequencies, indicating no departures from the simple asymptotic
expression.
In addition, it is likely
that the observed oscillation spectrum of 85~Peg~A will not show presence of
g modes nor of mixed modes. In fact, our theoretical calculation shows that
g modes are well confined in the core and with an amplitude which is too shallow
to be detected at the surface.
It is important to notice that the observed solar data at high frequencies
deviates from the model predictions. This is a consequence of a not fully
adequate treatment of the surface layers (see Sec. \ref{subsur}). 
Therefore this behaviour is also expected for
85~Peg~A which is, like the Sun, a low effective temperature star.
 
\section{Conclusions} 
 
In this paper we have addressed the problem of studying the structural 
properties and identifying the  evolutionary state of 85~Peg~A. The location in 
the H-R diagram indicates that this star is in the main-sequence phase of 
evolution,   with a mass in the range  \mass=0.75-0.82~\massun and with an 
age which can vary from 8 to 14~Gyr. These values have been obtained by 
assuming the most recent observational parameters of temperature and 
luminosity, but the classical method of fitting the stellar modeling parameters 
to the observational data does not allow to reduce the uncertainty on age and 
mass.  
 
Here we have shown that the problem of determining the  evolutionary state 
of 85~Peg~A will be solved once  observations of the oscillation spectrum 
will be available for this star. Theoretical calculation of models and p mode 
oscillations have shown that 85~Peg~A is a solar-like star, which is expected 
to oscillate in the range of frequencies $1000-5000\,\mu\mathrm{Hz}$. Our 
results show that its oscillation  spectrum seem to be characterized by a large 
separation for $l=0$ of  about $150-170\,\mu\mathrm{Hz}$ and a small separation 
between $l=0$ and $l=2$ which can 
vary from about $2$ to $6\,\mu\mathrm{Hz}$, according to the evolutionary 
state of the star. 
 
Indeed a difference of about 2~Gyr corresponds, in the asymptotic regime and 
hence at high frequencies, to a  difference in $\Delta_0$ of about 
$3\,\mu\mathrm{Hz}$ and in the ratio $d_{02}/\Delta_0$ of about $0.007$ (see 
Fig. \ref{f3}). It can be concluded that a determination of 
frequencies with an uncertainty $\leq 0.8\, \mu\mathrm{Hz}$, which corresponds 
to an average error of $\pm 0.005$  in the ratio  $d_{02}/\Delta_0$,  is 
sufficient to determine the age with an uncertainty better than 2~Gyr. 

Our work has examined the role of the most important physical inputs 
(metallicity, convection and diffusion processes) in the framework of the 
inputs of our evolutionary code. We have studied the influence of metal 
diffusion with the help of the CLES evolutionary code, which differs from ours 
in some physical inputs (e.g. nuclear reaction rates, treatment of helium 
diffusion, atmospheric boundary conditions, convection modelling). We therefore 
did not perform an accurate comparison of the codes, but simply have shown that 
the inclusion of metal diffusion has not an important impact on the age 
determination. Anyway, the change in metallicity implied by the \citet{thoul} 
description of metal diffusion  is relatively modest ($\sim 33$\%) in 12~Gyr, 
while FML02 find a reduction of metallicity by more than a factor 2 in $\sim 
9$~Gyr. This difference is taken by us as an indication that really much has to 
be done in the understanding of this issue.

We believe that this star should be considered as primary target for space or 
ground-based observations. It is clear indeed that only accurate 
observations will allow investigation of the  properties of the interior of 
this star and discrimination of its accurate evolutionary stage. 
The oscillation spectrum of 85~Peg~A will reveal how old is the star 
in terms of core hydrogen consumption, thanks to the very good mass constraint
coming from the astrometric and spectroscopic observations. 
In this respect, it appears feasible to put a lower limit to the age of the 
Galaxy based on the asteroseismology of 85~Peg~A.  

\section*{Acknowledgments} 

The authors are very grateful to G. Houdek for having contributed
to the paper with new computations
of the amplitude of the oscillations of 85 Peg A.
We are also grateful to the anonymous referee for helpful comments
and suggestions which led to the improvement of the present paper.
This project has been financially supported by Cofin 2002-2003 "Asteroseismology"
through the Italian Ministry of University and Research.

\label{lastpage} 
 
\end{document}